# AN EMPIRICAL INVESTIGATION ON SEARCH ENGINE AD DISCLOSURE

Dirk Lewandowski, Friederike Kerkmann, Sandra Rümmele, and Sebastian Sünkler

Hamburg University of Applied Sciences, Department of Information, Finkenau 35, D—22081 Hamburg, Germany

dirk.lewandowski@haw-hamburg.de, friederike.kerkmann@haw-hamburg.de, sa.ruemmele@gmail.com, sebastian.suenkler@haw-hamburg.de

## ABSTRACT

This representative study of German search engine users (N=1,000) focuses on the ability of users to distinguish between organic results and advertisements on Google results pages. We combine questions about Google's business with task-based studies in which users were asked to distinguish between ads and organic results in screenshots of results pages. We find that only a small percentage of users is able to reliably distinguish between ads and organic results, and that user knowledge of Google's business model is very limited. We conclude that ads are insufficiently labelled as such, and that many users may click on ads assuming that they are selecting organic results.

## KEYWORDS

Search engines, Google, search engine advertising, AdWords, representative study, questionnaire, online experiment

## INTRODUCTION

In recent years, there has been broad discussion of the role that Google plays in our society, and which role it should play, especially in European Union countries (Edelman & Lockwood, 2011; Manne & Wright, 2011; Thompson, 2011). The complaint that is best known among the public is being negotiated within the European Commission's antitrust investigation of Google. The probe was initiated in 2010, but no resolution has yet been achieved (European Commission, 2015). The reason Google's role is so important derives from its huge market share in the majority of global markets. For instance, in the U.S., Google accounts for roughly two thirds of the search market ("Stats: comScore," 2015), while in Europe, it exerts even greater dominance with a market share of well over 90% in nearly every country (European Commission, 2015).

Search engines generate their revenues through advertising, and search engine advertising has become a multi-billion-Euro industry (Knapp & Marouli, 2015). The basic assumption of search engine advertising is that a user searching for something already reveals his or her intent with the search query (Battelle, 2005, p. 1). Advertisements can then be placed, not disturbing the user, but helping to find a product or service he or she is actually looking for. In this sense, these ads can be regarded as a type of search result.

Since advertisers do not pay when their ads are displayed, but rather when they are clicked on (Danescu-Niculescu-Mizil, Broder, Gabrilovich, Josifovski, & Pang, 2010, p. 292), it is in the interest of search engine vendors to generate the highest possible number of ad clicks. So search engine vendors may be tempted to blur the line between paid advertising and organic results to make more money. An initial indicator can be the design and structure of the advertisements if they are very similar to organic results (Figure 1). In the case of Google, both contain a title, a short description ("snippet") and a URL. Also, the colour scheme used is the same: the title is blue and underlined on mouse-over, the description black, and the URL is also blue. If one were to see a description of an organic result and an advertisement out of context, it would be difficult to distinguish between the two. This leads to the central question of our research, namely *whether search engine users are able to distinguish between paid advertising and organic results on the search engine results pages*.

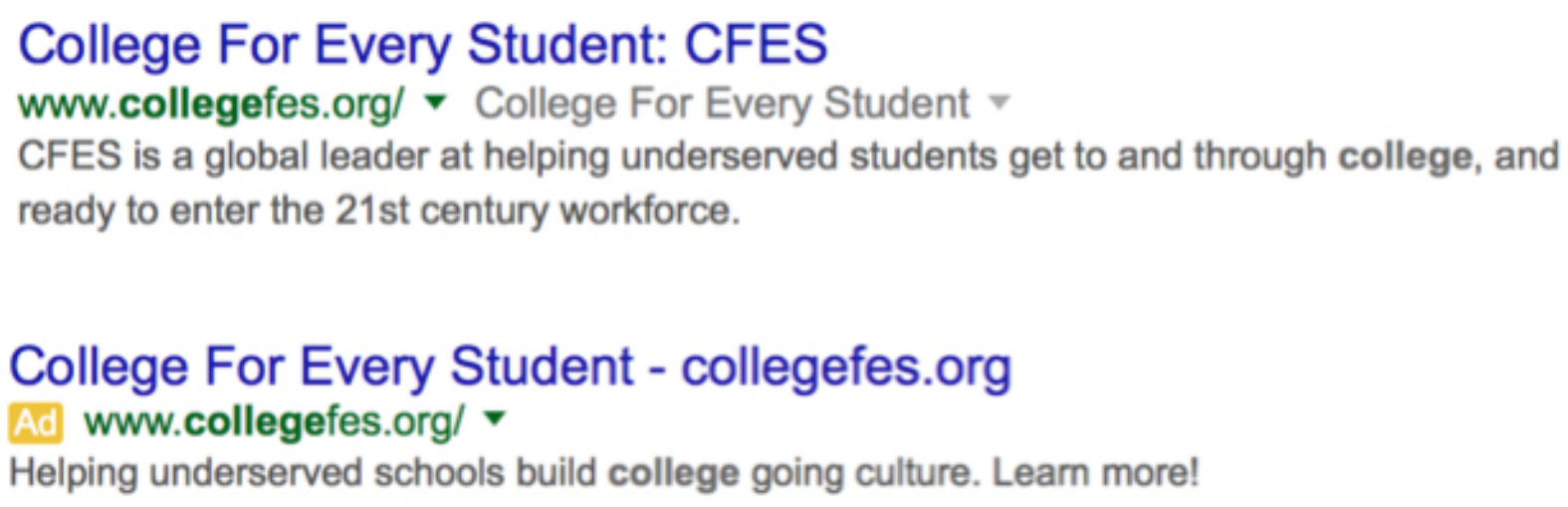

***Figure 1*. a) organic result, b) paid result (both from Google)**

## LITERATURE REVIEW

### Search engine results pages

Search engines present different types of results on their search engine results pages (SERPs). A search engine *results page* is the complete HTML page output that a search engine serves in response to a search query entered by a user. In contrast, a *results screen* is the part of the SERP that a user can see without scrolling down.

Results presentation on SERPs has changed in recent years.[1] The simplest model of a SERP is a ranked list of documents provided in response to a query. However, once search engines started displaying ads on SERPs, there were actually two ranked lists: the list of "organic results" and the list of advertisements. By adding results from vertical search engines such as news and integrating them into the SERPs (known as *Universal Search*, see Taylor, Mayer, & Buyukkokten, 2008), search engines moved away from plain ranked results lists and on to a richer presentation both of individual results as well as certain results types. Current SERPs go even further to additionally display factual information. These are shown in what are known as *Knowledge Graph* results (Drumond Monteiro & Aparecida Moura, 2014), satisfying at least some information needs directly on the results pages and representing a departure from the concept of a search engine being a tool for sending traffic (i.e., users) to external web pages.

Earlier studies defined relevant areas of SERPs either on a micro-level (i.e., differentiating the type of snippets shown on the SERPs; in Höchstötter & Lewandowski, 2009) or on a macro-level (i.e., distinguishing between organic results and advertising space; Nicholson et al., 2006). In the current study, we follow this macro-level distinction, but add two other types of results, namely Universal Search results and Knowledge Graph results.

---

[1] In the context of the present study, we discuss results pages on the desktop versions of search engines.

We define the four areas of SERPs as follows:

- *Organic results* are results that are generated from the search engine's index of web pages. Every document in that index has the same chance of being displayed in response to a certain query, as all documents are treated the same by the ranking algorithms.
- *Advertisements* in the context of search engines are text-based. They are also shown as a response to a query, and form a separate results list (or more than one separate results list) on the SERP.
- *Universal Search results* are results generated from vertical search engine indexes, such as news or images. Depending on the nature of the index, these results can either be generated similarly to organic results (as in the case of images) or be based on a certain collection of sources (as in the case of news, where a collection of trusted news sources is defined beforehand by the search engine vendor). Universal Search results can also come from document collections especially built by the search engine vendor (as opposed to the results from the web index that come from a multitude of sources distributed across the web).
- *Knowledge-graph results* are results within which the search engine displays actual answers or facts instead presenting links on the SERPs.

Some studies have investigated user viewing behaviour on SERPs using eye-tracking. They found that, for list-based presentations (i.e. results are presented in the form of a single ranked list), users strongly prefer the first few results, and that the time spent on the results descriptions ("snippets") decreased heavily after the first result (Granka, Joachims, & Gay, 2004). In more complex results presentations like Universal Search, these patterns change (Liu, Liu, Zhou, Zhang, & Ma, 2015). Still, the ranked list has an influence on what elements a user perceives, but the different designs of certain results types (like images already shown in the results list) guide the user towards certain elements on the results screen. Therefore, reading behaviour on the SERPs differs heavily from one SERP to the next, depending on which elements are shown at which position.

When looking at how users *select* results from a SERP, they are influenced mainly by the following factors:

1. Results position and reading behaviour: users tend to click on results at or near the top of a results list (Joachims, Granka, Pan, Hembrooke, & Gay, 2005). For instance, a 2014 study from the software company Caphyon used 465,000 queries and analysed the click-through rates in Google. They found that more than two thirds of all clicks go to the first five positions, and the result ranked first accounts by itself for 31% of all clicks (Petrescu, 2014). Goel et al. (2010) found that within Yahoo search, only 10,000 websites account for approximately 80% of results clicks. This clicking behaviour is due to the fact that, usually, lists of results are read from top to bottom, so that users pay little or no attention to items shown further below.
2. Search engine relevance ranking algorithms are precision-based, i.e., they focus on presenting a few relevant results in the first several positions. Users have adapted to this kind of results ranking and therefore in most cases only consider the first few results.
3. Due to screen resolutions and browser window sizes, SERPs are divided into an area "above the fold" and an area "below the fold" (B. J. Jansen & Spink, 2007). Users predominantly click on results shown above the fold.

Nicholson et al. (2006) found that only 40% of the results on the first screen (i.e., the area visible to the searcher without scrolling down) were organic results. This ratio increased to 67% when considering the first results page. This study, however, is limited by the low number of queries used and the fact that only one (rather low) screen resolution was considered.

### Trust in search engine results

People tend to trust information they find on the Internet, and do so even more when it is presented in search engine results, especially when those results are from Google. As early as 2003, Graham & Metaxas (2003) summarized their finding from a user study with a quote from one of the participants: "Of course it's true, I saw it on the internet". Users often see Google as a source rather than an intermediary between a user and multiple sources. Further studies have meanwhile confirmed these findings (see Tremel (2010).

Users assign trust in the retrieved documents based on the search engine rather than (or in addition to) the actual source. This can also be seen in user "information repertoires" (i.e., the sources users obtain information about current events from etc. (see Hölig & Hasebrink, 2013) — participants listed not only printed and online news sources as being relevant for news consumption, but also Google.

While search engine rankings can only "simulate" concepts such as trust and credibility (Lewandowski, 2012) in that they use technical factors for assuming whether a document is trustworthy/credible, users tend to see search engine rankings as trustworthy. Pan et al. (2007) found that users trusted Google's results ranking even more than their own judgments when it comes to choosing relevant results from the ranked list. A representative study of U.S. internet users, (Purcell, Brenner, & Raine, 2012) found that "91% of search engine users say they always or most of the time find the information they are seeking when they use search engines, 73% of search engine users say that most or all the information they find as they use search engines is accurate and trustworthy, 66% of search engine users say search engines are a fair and unbiased sources of information." These findings suggest that there is a high correlation between users' confidence in their successful use of search engines and their comprehension of search engines as being purely technical/algorithmic systems in which issues of trust and bias are not relevant.

### Ad labelling

It is important to note that contextual, text-based ads ("sponsored links", "paid results") can be seen as one type of search result. As said in the introduction, both results types are constructed and designed in a similar way. Therefore, it seems reasonable to suppose that users may find it difficult to distinguish between the two results types. Ads may be relevant to a query. Contrary to other forms of advertising, search-based ads target a user who has an explicit interest in something. Jansen (2007) found that for e-commerce related queries, the relevance ratings for organic results and ads are practically the same. It should be stressed that ads *can* be helpful to satisfy a user's information need, and therefore, the distinction between organic results and ads may not be important to users seeking information in areas where ads can serve as pointers to content that satisfies their information needs.

Considering the four areas of the SERPs discussed above, it may be hard for users to distinguish between elements which were produced on the basis of objective criteria and elements which are produced on the basis of commercial decisions. When considering the different results types, especially organic results and Universal Search results, there may be blurred lines between paid and unpaid results, especially when Universal Search results point to offerings by the search engine used. An early study (Fallows, 2005) found that only 38% of U.S. searchers were aware of the distinction between organic and paid results. The situation has surely changed since then, partly due to the U.S. Federal Trade Commission's guidelines on search engine ad disclosure (Sullivan, 2013a). However, the distinction between the two results types is still an issue, not only in general-purpose Web search engines, but also in many specialized vertical search engines (Sullivan, 2013b). Some industry studies strongly suggest that the labelling of ads might not be clear enough (Bundesverband Digitale Wirtschaft, 2009; Charlton, 2013; Wall, 2012).

### Search engine impact on knowledge acquisition

As is well known, search engines are a major means of acquiring information from the Web. They are important based simply on the number of queries they process. Google alone serves more than 1 trillion searches a year (Sullivan, 2015). For each query, the SERP presents a certain selection and order of results, which influences what is finally clicked on. As each result set is the product of a certain interpretation of the contents available on the Web, a search engine with a market share like Google's not only dominates the search market in terms of queries answered or advertising revenues, but also in terms of presenting results that conform to the search engine vendor's beliefs and assumptions, as expressed in the algorithms designed by its employees (Mager, 2012; van Couvering, 2007). Users choose the information they consume through search engines from a very limited set of results, and they are also influenced by the way these results are presented. While search engines have huge indexes of web pages and make it possible to find very specific information from a huge variety of sources, in practice, there is a heavy bias towards some popular websites (Goel, Broder, Gabrilovich, & Pang, 2010; Höchstötter & Lewandowski, 2009).

Google's market share represents a huge influence on what users get to see (and to choose from) when they use a search engine. This leads to the question of whether search engines providers, especially when they have a huge market share, have some responsibility to inform users on how results rankings are generated and in which way their results may be influenced by factors other than quality-based ranking functions.

In summary, we see that in the design of the search engine results pages (SERPs), search engine companies make choices that affect the way these pages are perceived by the users, following certain viewing and selection behaviour. While users trust search engines in that they provide them with the most relevant results on the first positions, findings on ads labelling suggest that results presentation may be too complex for users to reliably distinguish between content that has been paid for and so-called organic content. As search engines play a central role for not only accessing content on the web, the design of SERPs in general, and the labelling of advertisements in particular, have a huge impact on what kind of information users actually see in response to their queries.

## OBJECTIVES AND RESEARCH QUESTIONS

The objective of this study is to determine whether users are able to distinguish between organic results and advertising on Google's results pages. We approached this objective by designing a study that combines user self-assessment of knowledge about search engines and Google's revenue model in particular, and a click-based test where users had to label advertising and organic results, respectively.

Our research questions are as follows:

**RQ1: Are users able to distinguish between organic results and ads on Google's search engine results pages?**

Distinguishing between ads and organic results is a prerequisite for making informed choices in results selection. Searchers to a vast majority trust in search engine results (Purcell et al., 2012) and given that they are not aware of clicking on ads, they "are more likely to encounter faulty or biased information on Web pages of companies that can afford to be listed on the first results page, yet do not necessarily have the most accurate or unbiased information." (Marable, 2003) With this research question, we aim to get knowledge on how many users are actually affected by the potential problem with distinguishing the results types. The answer will give supporting information on potential regulatory steps.

To look further into differences between user groups, we added the following three research questions, segmenting the sample.

**RQ1a: Are there age differences when it comes to distinguishing organic results and ads?**

It is often proclaimed that younger users are better able to use search engines. Studies reporting on the different aspects of searching behaviour come to varied results when comparing age groups (see (Singer, Norbisrath, & Lewandowski, 2012, p. 25ff. for an overview). Potential problems with not understanding the composition of SERPs on the one hand may be seen as a matter of time only, as younger and more competent users will grow up and replace older users. However, if there are no age differences found in that respect, this would call for more information literacy instruction for young users.

**RQ1b: Are there differences regarding the level of education when it comes to distinguishing organic results and ads?**

The Web is used differently by users of different levels of education (ARD/ZDF, 2015). It may therefore also be the case that distinguishing between ads and organic results is merely a problem with certain educational groups. From an information literacy point of view, this would call for addressing these groups individually.

**RQ1c: Are there differences regarding self-assessed searching skills when it comes to distinguishing organic results and ads?**

The majority of users state that they are well able to find what they are looking for when using search engines (Purcell et al., 2012). However, user studies found that many users lack competence in formulating precise queries (e.g., (Höchstötter & Koch, 2009; Stark, Magin, & Jürgens, 2014) and in judging the trustworthiness of results, as well (Stark et al., 2014). Therefore, it is important to find whether self-reported measures on searching skills correlate with task-based measures.

**RQ2: How do users think search engines make money?**

Only when users understand that advertising forms the basis of search engines' business models (J. Jansen, 2011), they can be aware that on the SERPs, not everything is organic results (cf. (Höchstötter & Lewandowski, 2009). With this research question, we want to find out what ideas users actually have on how search engine companies earn their revenues, and quantify these data.

## METHODS

We used a mixed approach, containing an online questionnaire, as well as some clickable and markable screenshots that were sent to 1,000 German internet users. The sample is representative of the German online population, as according to the criteria applied by AGOF, a leading German online panel ("Method - AGOF coverage currency," 2015). AGOF provides a standardised online coverage currency to measure the success of marketing tools. The currency is based on a Three-Pillar Model for Data mining and profiling by electronic measurement of page visits and page impressions, by on-site surveys on descriptive socio-demographic values and representative telephone surveys. The population includes Internet Users from the age of 10 years. The sampling is based on recruiting participants from different slices of the sample. They are invited until the desired number of participants in each slice is reached. Due to the AGOF procedures, we are unfortunately not able to give numbers on how many users were asked to participate but declined. A market research firm carried out the survey in December 2013.

The questionnaire contained two parts:

1. In the first, we asked users about their knowledge of advertisements in search engines, and about their knowledge of the search engine Google.
2. In the second part, users were given tasks where they had to either mark ads or organic results on screenshots of SERPs.

As search results presentation has changed a lot in recent years, we tried to at least address the major elements of results presentation in our study. We sought good combinations of these elements in order to also identify problems in distinguishing between different results elements. In addition to the different results types, we also tried to cover different query intents. We covered informational as well as transactional queries (cf. Broder, 2002), and also commercial intent (cf. (Lewandowski, Drechsler, & Mach, 2012). A third criterion was whether the SERP contained Universal Search results or not. Again, as it was not possible to cover all possible combinations, we tried to cover some typical combinations of the criteria.

When users logged in into the questionnaire, they were first asked whether they had used the search engine Google within the last three months (question 1). All participants were Google users, so no participants were excluded.

Then, participants answered some questions about their knowledge of search engines and Google's business (question 2-6) before being asked in the main part of the questionnaire to mark advertisements and organic results, respectively, on screenshots of SERPs (question 7-11).

## RESULTS

### Characteristics of the participants

With the question 2, we asked the participants to rate their competency when it comes to searching in Google. We used the German grading system, where 1 is the best grade, and 6 the worst. The vast majority (90.8%) rated themselves as either "very good" (grade 1) or "good" (grade 2). There were only a few users who rated their competency with a 4 or worse.

### User knowledge of Google's business

Next, we asked participants in an open question to describe in their own words how Google generates its revenue. They filled in their own answers, which were later classified by a research assistant. 81% correctly named advertising as the source of Google's revenues. However, only 60.6% named advertising as Google's *only* source of revenue, and 20.4% mentioned other, incorrect sources of revenue. 9.5% gave an outright wrong answer, and another 9.5% said they did not know.

In the next question, we asked users whether it was possible to pay Google for giving one's company a preferred listing on the search results pages for a particular search query. With the wording of this question, we wanted to make sure that, hypothetically, both paid placements in organic results and placements of contextual advertisements were covered.

73.3% of participants correctly said that such a listing was possible. 6.4% said this was not possible, while 20.3% said they didn't know.

We then asked the users answering the last question with "yes" whether it was possible to distinguish between paid advertisements and unpaid results on Google's SERPs. 57.98% said this was possible, while 26.6% said it was not, and 15.42% said they didn't know.

Taken the last two questions together, we can see that 42% of the German internet users self-report that they either do not know that it is possible to pay Google for preferred listings for one’s company on the SERPs, or do not know how to distinguish between organic results and ads.

With question 6, we asked participants who said that it was possible to distinguish between paid advertisements and unpaid results, in which way these two results types are differentiated. This again was an open question where we asked the participants to name the most important differences. Again, a research assistant classified the answers into categories. Note that it was possible to mention more than one option.

There were only 37 participants who named elements distinguishing ads from organic results correctly, namely:

1. Shading and/or different layout
2. Shown at the top of the results page, before the organic results
3. Shown on the right-hand side of the SERP

Many users named some of the correct elements, but not the complete set.

As a summary of our questions on users’ self-reported knowledge about advertisements in the Google search engine, we can say that a large portion of Google users is not able to distinguish between organic results and advertisements. In Figure 2, the segmentation of users from questions 4 to 6 is summarized. In total, 62.2% of participants were not able to distinguish between ads and organic results. They either said it was not possible to pay Google for a preferred listing of one’s company on the SERPs, that they did not know if this was possible, or they said that it was not possible to distinguish between ads and organic results (or they did not know), or they named incorrect ad labelling (or did not know).

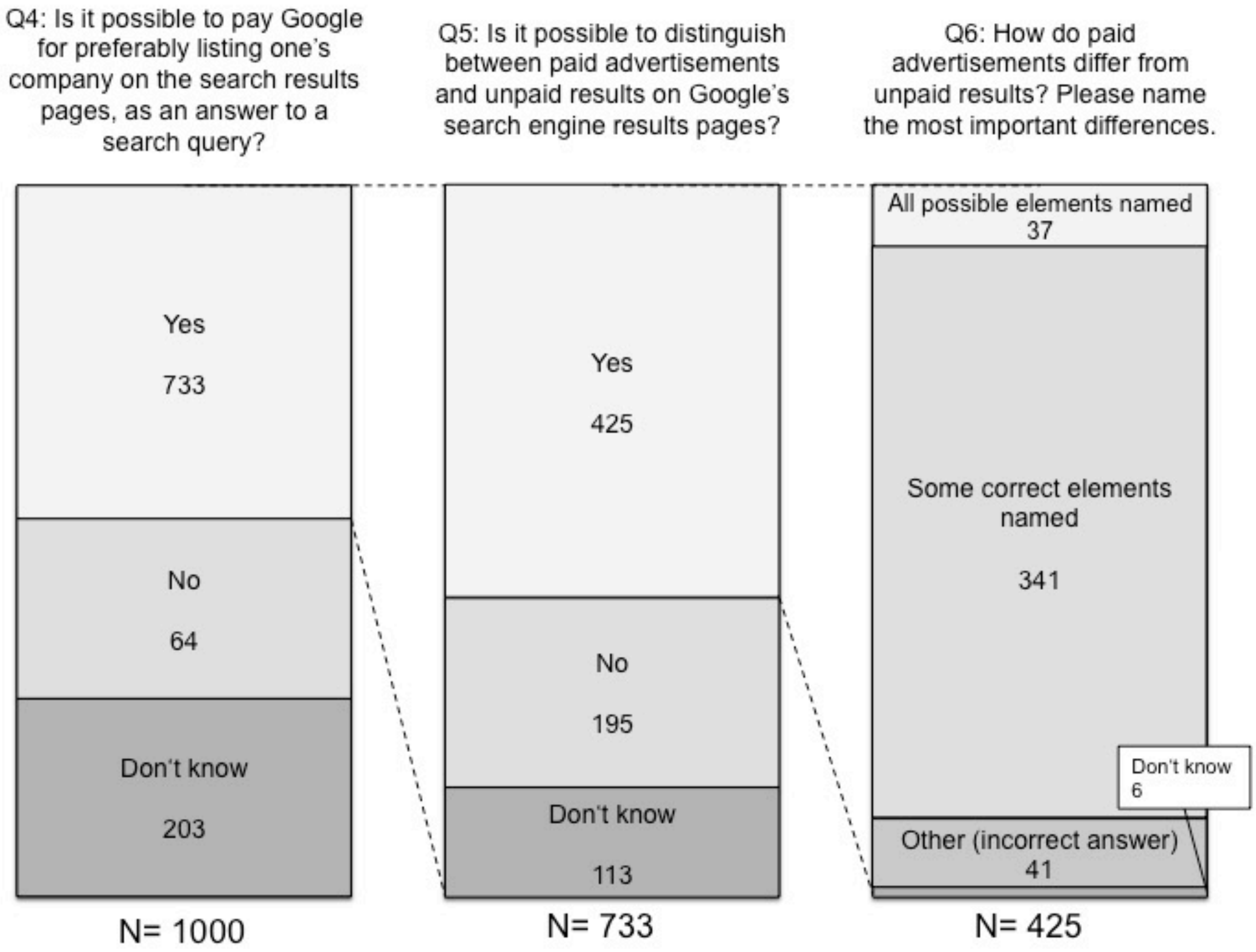

***Figure 2*. Summary of the findings on users' knowledge of Google's business**

**Screenshot-based tasks**

In the screenshot-based tasks, participants were asked either to label all ads on a screenshot of a SERP or label all organic content on that page. In total, participants were given five such tasks, which allows for general conclusions to be drawn to an extent across all the tasks. Figure 3 shows how participants labelled the areas on the screenshots using frames. They were able to select entire areas to label the results. We defined pixel areas on the screenshots for evaluating the selection of the participants to the given tasks.

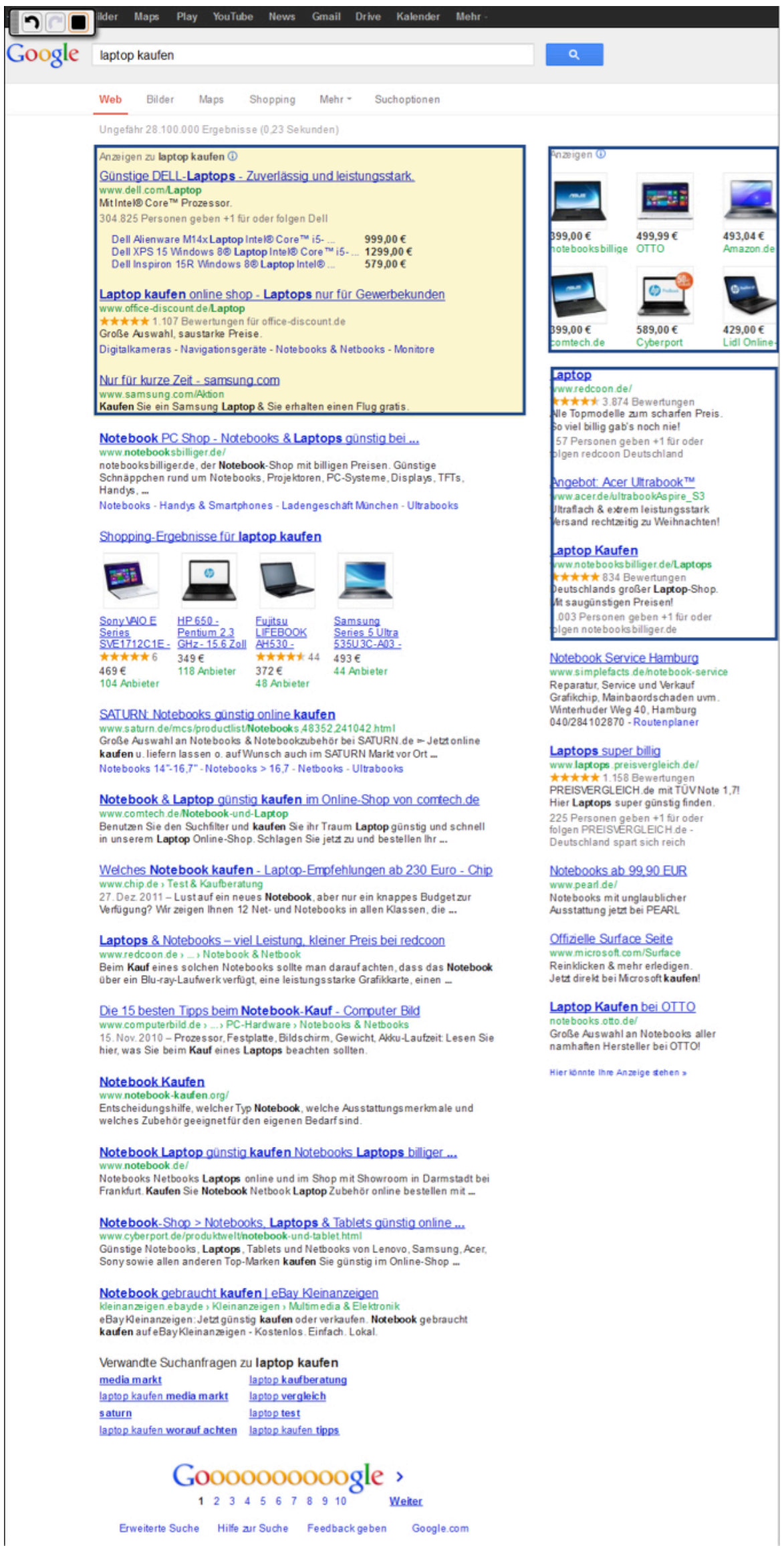

***Figure 3.* Labelling of certain areas on a screenshot**

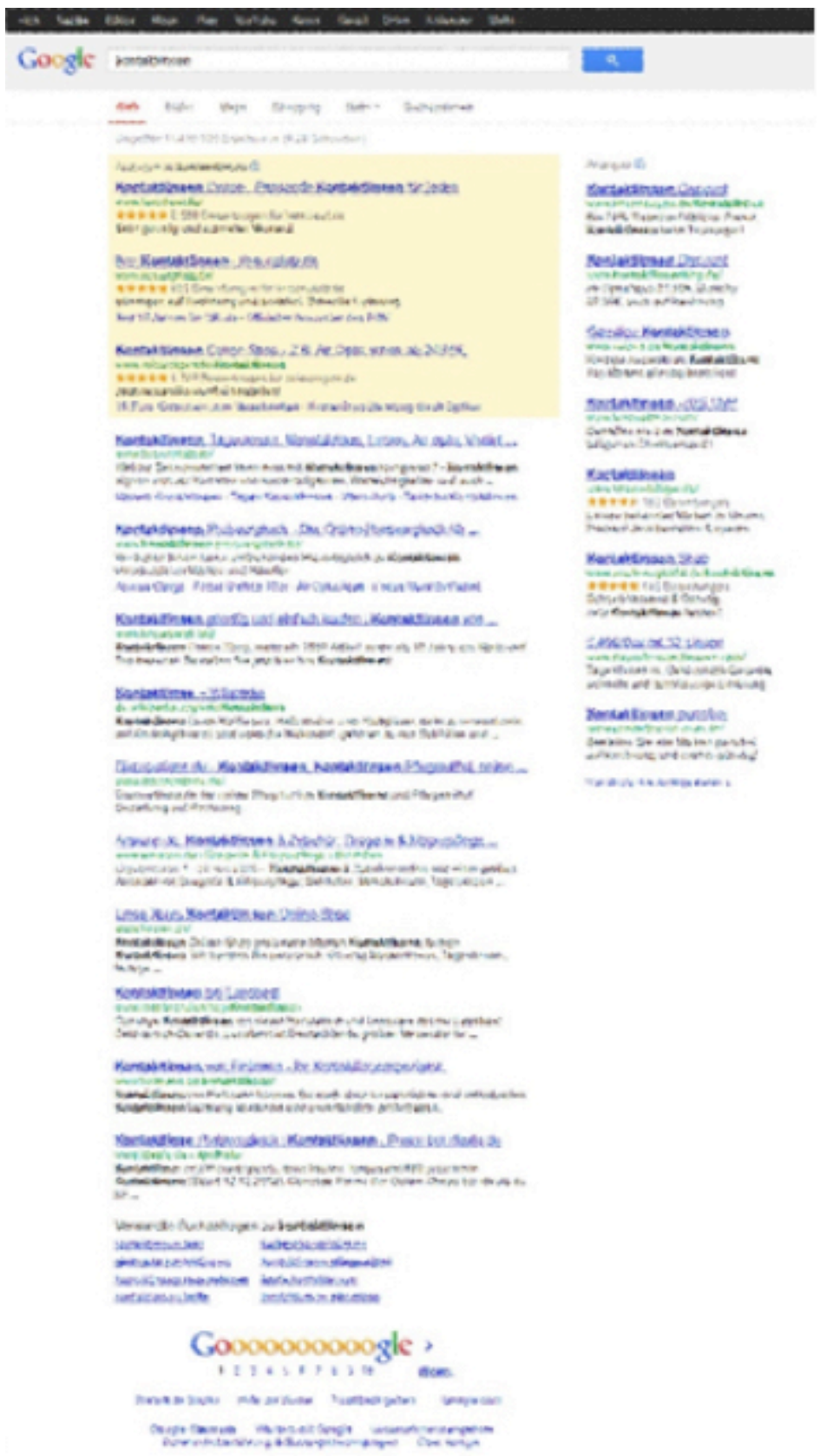

*Figure 4.* **Pixelated screenshot for Q7**

In the first question using screenshots (question 7), we pixelated the SERP with paid and unpaid results to find out if users were able to distinguish between organic results and advertisements solely on the basis of the structure of the SERP (see Figure 4).

We found that only 25.8% of the users were able to mark everything correctly (see table 1). So we can see that approximately three quarters of the users are not able to reliably recognize advertisements. 63.9% of participants marked only ads, i.e., they did recognize some advertisements as ads, but did not label *all* advertisements.

10.9% marked only ads on the right-hand side of the SERP, but not the ads shown above the list of organic results. On the other hand, 4.6% only recognized ads above the organic results, but not on the right-hand side of the SERP.

12.4% of participants labelled at least one organic result as an advertisement, an additional 0.6% marked all results shown as ads. Taken together, 13% of the users are not able to distinguish between the two results types at all.

In the following analyses, we used the analysis of variance (ANOVA) test for testing for statistical differences between groups (grades, knowledge on Google's business, etc.). There were no significant differences between the users groups based on their self-assessments regarding their searching skills (see table 1). However, there were differences between users who knew that Google is financed through advertising, and users who did not (see table 2). It is obvious that users who know that there is advertising on the SERPs are better able to identify ads. Users who named other forms of income in addition to ads were, however, also better able to label the ads than users in the group who did not answer the question on Google's revenue model correctly, as were those who claimed they did not know the answer.

There were significant differences between the group that knew that it is possible to place ads on the SERPs, the group that answered that question incorrectly, and the group answering "I don't know". Again, this last group performed worst on the task. Nevertheless even the best-performing group could only label the results correctly approximately 30% of the time.

**Table 1.** ***Results for pixelated screenshot (question 7) by self-assessments of the participants regarding searching skills***

| | All users | German grade 1 (n = 454) | German grade 2 (n = 454) | German grade 3 (n = 81) | German grade 4 (n = 9) | German grade 5 (n = 1) | German grade 6 (n = 1) |
|---|---|---|---|---|---|---|---|
| | | % | % | % | % | % | % |
| ALL areas marked correctly | 25.8 | 25.6 | 27.1 | 19.8 | 33.3 | 0.0 | 0.0 |
| Advertisements only (but did not identify all advertisements) | 63.9 | 63.4 | 64.5 | 64.2 | 66.7 | 0.0 | 0.0 |
| Organic results marked as advertisements | 12.4 | 11.9 | 12.1 | 16.0 | 11.1 | 100.0 | 0.0 |
| All results shown marked as ads | 0.6 | 0.2 | 0.9 | 1.2 | 0.0 | 0.0 | 0.0 |
| Marked only ads on the right-hand side | 10.9 | 9.5 | 12.3 | 12.3 | 0.0 | 0.0 | 0.0 |
| Marked only ads above the organic results | 4.6 | 5.5 | 3.7 | 4.9 | 0.0 | 0.0 | 0.0 |

*Note.* * p < 0.05. For German grades, 1 is the best possible grade and 6 the worst.

**Table 2.** ***Results for pixelated screenshot (question 7) by self-reported knowledge about Google's business model; knowledge of placing ads on the SERPs; and knowledge of the distinction between advertisements and organic results***

| | Knowledge of Google business model | | | | Knowledge of paying Google to place ads on the SERPs | | | Knowledge of distinction between advertisements and organic results** | | |
|---|---|---|---|---|---|---|---|---|---|---|
| | correct (n = 606) | incorrect (n = 93) | don't know (n = 95) | incorrect statement (n = | yes (n = 733) | no (n = 64) | don't know (n = 203) | yes (n = 425) | no (n = 195) | don't know (n = 113) % |
| | % | % | % | 206) % | % | % | % | % | % | |
| ALL areas marked correctly | 28.1 * | 23.7 * | 10.5 * | 27.2 * | 30.0 * | 21.9 * | 11.8 * | 36.7 * | 19.5 * | 23.0 * |
| Advertisements only (but did not identify all advertisements) | 67.2 * | 63.4 * | 47.4 * | 62.1 * | 65.6 | 57.8 | 59.6 | 70.6 * | 56.4 * | 62.8 * |
| Organic results marked as advertisements | 12.9 | 10.8 | 13.7 | 11.2 | 13.0 | 12.5 | 10.3 | 11.3 | 16.4 | 13.3 |
| All results shown marked as ads | 0.7 | 1.1 | 0.0 | 0.5 | 0.5 | 0.0 | 1.0 | 0.2 | 1.0 | 0.9 |
| Marked only ads on the right-hand side | 12.0 | 9.7 | 11.6 | 7.8 | 10.1 | 6.3 | 15.3 | 8.7 * | 7.7 * | 19.5 * |
| Marked only ads above the organic results | 4.6 | 4.3 | 4.2 | 4.9 | 5.0 | 1.6 | 3.9 | 5.9 | 4.6 | 2.7 |

*Note.* * $p < 0.05$. ** *FILTER: only participants who answered the former question with "yes". (n=733)*

In question 8, participants were shown a non-pixelated results page containing organic results and advertisements (Figure 5). We asked them to mark every single advertisement. 35% were able to mark all ads correctly (without marking any organic results), while 18% marked at least one organic result as an advertisement. Again, some users marked only the ads on the right-hand side or the ads shown at the top of the results list as advertisements (5.9% and 6.1%, respectively).

Regarding self-reported search knowledge, there were no significant differences (see table 3). However, participants who performed better on the questions regarding Google's business also performed better on labelling the ads in this question (table 4).

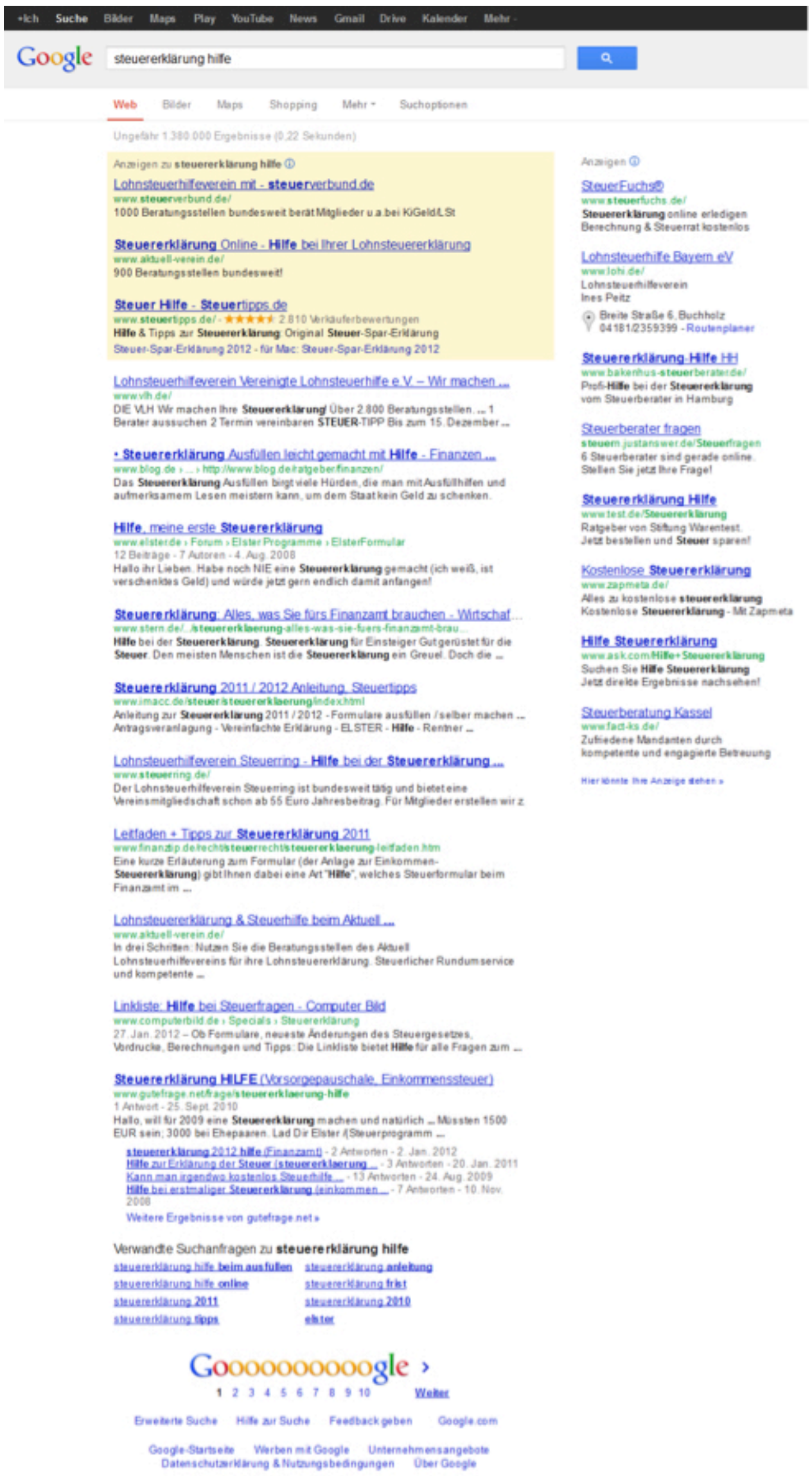

***Figure 5.*** **Screenshot provided for the taxes task (question 8)**

**Table 3.** ***Results for taxes task (question 8) by self-assessments of the participants regarding search skills***

| | All users | German grade 1 (n = 454) | German grade 2 (n = 454) | German grade 3 (n = 81) | German grade 4 (n = 9) | German grade 5 (n = 1) | German grade 6 (n = 1) |
|---|---|---|---|---|---|---|---|
| | | % | % | % | % | % | % |
| ALL areas marked correctly | 35.0 | 33.5 | 36.6 | 33.3 | 44.4 | 0.0 | 100.0 |
| Advertisements only (but did identify all advertisements) | 66.3 | 66.7 | 67.2 | 58.0 | 66.7 | 100.0 | 100.0 |
| Organic results marked as | 18.0 | 17.6 | 16.5 | 28.4 | 22.2 | 0.0 | 0.0 |
| advertisements | | | | | | | |
| All results shown marked as ads | 0.7 | 0.2 | 1.1 | 1.2 | 0.0 | 0.0 | 0.0 |
| Marked only ads on the right-hand side | 5.9 | 5.5 | 6.6 | 4.9 | 0.0 | 0.0 | 0.0 |
| Marked only ads above the organic results | 6.1 | 33.5 | 36.6 | 33.3 | 44.4 | 0.0 | 100.0 |

*Note.* * $p < 0.05$. For German grades, 1 is the best possible grade and 6 the worst.

**Table 4.** ***Results for taxes task (question 8) by self-reported knowledge about Google's business model by knowledge of placing ads on the SERPs and by knowledge of distinction between advertisements and organic results***

| | Knowledge of Google's business model | | | | Knowledge of paying Google to place ads on the SERPs | | | Knowledge of distinction between advertisements and organic results** | | |
|---|---|---|---|---|---|---|---|---|---|---|
| | correct (n = 606) | incorrect (n = 93) | don't know (n = 95) | incorrect statement (n = 206) | yes (n = 733) | no (n = 64) | don't know (n = 203) | yes (n = 425) | no (n = 195) | don't know (n = 113) |
| | % | % | % | % | % | % | % | % | % | % |
| ALL areas marked correctly | 36.6 * | 28.0 * | 14.7 * | 42.7 * | 38.7 * | 39.1 * | 20.2 * | 47.1 * | 25.1 * | 31.0 * |
| Advertisements only (but did not identify all advertisements) | 70.0 * | 53.8 * | 47.4 * | 69.9 * | 69.0 * | 70.3 * | 55.2 * | 76.7 * | 52.8 * | 68.1 * |
| Organic results marked as advertisement | 16.7 | 24.7 | 17.9 | 18.9 | 18.4 * | 6.3 * | 20.2 * | 12.7 * | 29.2 * | 21.2 * |
| All results shown marked as ads | 0.5 | 2.2 | 1.1 | 0.5 | 0.5 | 0.0 | 1.5 | 0.5 | 1.0 | 0.5 |
| Marked only ads on the right-hand side | 6.8 | 4.3 | 3.2 | 5.3 | 5.6 | 3.1 | 7.9 | 4.9 | 7.7 | 4.4 |
| Marked only ads above the | 5.9 | 5.4 | 7.4 | 6.3 | 5.6 | 10.9 | 6.4 | 6.1 | 3.1 | 8.0 |

| |
|---|
| organic results |

*Note.* *$p < 0.05$, ** *FILTER: only participants answered the former question with "yes". (n=733)*

The next screenshot again showed a SERP with organic results and ads both on top of the results and on the right-hand side (figure 6). The task was to buy glasses. In this case, participants were asked to mark all organic results. Only 19% marked all organic results correctly without marking any ads. 27.6% marked at least one ad as an organic result, and 32.3% only marked organic results, but not the complete set (Table 5). There were no significant differences regarding the users' self-reported search knowledge. We found significant differences for users knowing about the practices of placing ads on the SERPs and about how to distinguish between ads and organic results (see Table 6).

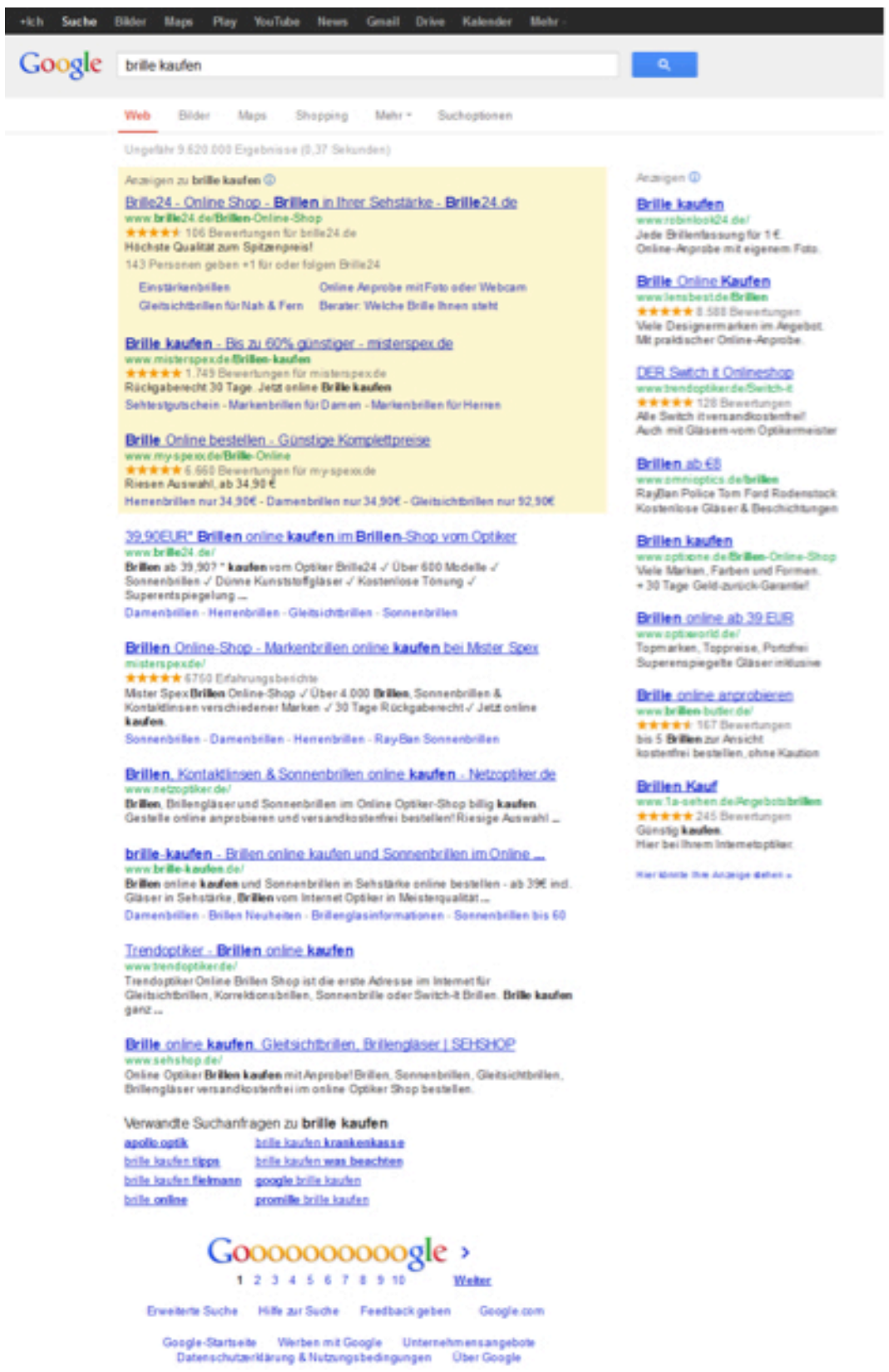

***Figure 6*. Screenshot provided for the "buy glasses" task (Q9)**

**Table 5. *Results for "buy glasses" task (question 9) by participants self-assessment regarding search skills***

| | All users | German grade 1 (n = 454) | German grade 2 (n = 454) | German grade 3 (n = 81) | German grade 4 (n = 9) | German grade 5 (n = 1) | German grade 6 (n = 1) |
|---|---|---|---|---|---|---|---|
| | | % | % | % | % | % | % |
| ALL organic results marked correctly | 19.0 | 16.3 | 21.4 | 22.2 | 11.1 | 0.0 | 0.0 |
| Organic results only (but did not mark all organic results correctly) | 32.3 | 30.0 | 34.6 | 33.3 | 22.2 | 100.0 | 0.0 |
| Advertisements marked as organic results | 27.6 | 28.4 | 28.4 | 18.5 | 22.2 | 0.0 | 100.0 |
| All results shown marked as organic results | 1.6 | 1.3 | 2.0 | 1.2 | 0.0 | 0.0 | 0.0 |

*Note.* * $p < 0.05$. For German grades, 1 is the best possible grade and 6 the worst.

**Table 6.** ***Results for "buy glasses" task (question 9) by self-reported knowledge about the Google business model by knowledge of placing ads on the SERPs and by knowledge of distinction between advertisements and organic results***

| | Knowledge of Google's business model | | | | Knowledge of paying Google to place ads on the SERPs | | | Knowledge of distinction between advertisements and organic results** | | |
|---|---|---|---|---|---|---|---|---|---|---|
| | correct (n = 606) | incorrect (n = 93) | don't know (n = 95) | incorrect statement (n = 206) | yes (n = 733) | no (n = 64) | don't know (n = 203) | yes (n = 425) | no (n = 195) | don't know (n = 113) |
| | % | % | % | % | % | % | % | % | % | % |
| ALL organic results marked correctly | 20.3 | 15.1 | 12.6 | 19.9 | 21.3 * | 10.9 * | 13.3 * | 26.8 * | 14.4 * | 12.4 * |
| Organic results only (but did not mark all organic results correctly) | 34.0 | 28.0 | 26.3 | 32.0 | 35.2 * | 21.9 * | 25.1 * | 39.5 * | 27.7 * | 31.9 * |
| Advertisements marked as organic results | 28.7 | 24.7 | 22.1 | 28.2 | 27.0 | 34.4 | 27.6 | 26.6 | 25.1 | 31.9 |
| All results shown marked as organic results | 2.0 | 1.1 | 1.1 | 1.0 | 1.4 | 1.6 | 2.5 | 0.9 | 2.1 | 1.8 |

*Note.* * $p < 0.05$, ** *FILTER: only participants answered the former question with "yes". (n=733)*

In question 10, we showed participants a screenshot of a SERP for the query "Vivienne Westwood" and asked them to mark any advertisements on the page. In this case, the correct solution was to mark advertisements on top of the results list, at the bottom of the results list, and the ad (showing a shoe) on the right-hand side (below Knowledge Graph results; see Figure 7).

15.8% were able to label all advertisements correctly, while 44.9% marked only advertisements, but not necessarily the complete set. 31.9% marked at least one organic result as an ad, and 14.6% marked a Universal Search result (Google Shopping result or Knowledge Graph result) as an ad (see Table 7). It should be noted that at the time when the study was conducted, Google Shopping results were not ads, and were only later changed to "Google Product Ads" (Marvin, 2014).

As seen in table 7 there were significant differences with users' self-reported search knowledge regarding Universal Search results and ads on the right hand side: Users with worse self-reported searching knowledge were more likely to mark Universal Search results as advertisements, and also more likely to only mark ads on the right-hand side instead of the full set of advertisements.

We also found significant differences regarding user knowledge of Google's revenues: users who have better knowledge are more likely to mark all ads correctly, and also to mark only advertisements (regardless of whether the whole set of ads is marked or only some ads). Also, users who say that ads and organic results are distinguishable on the SERPs perform better concerning all variants in table 8.

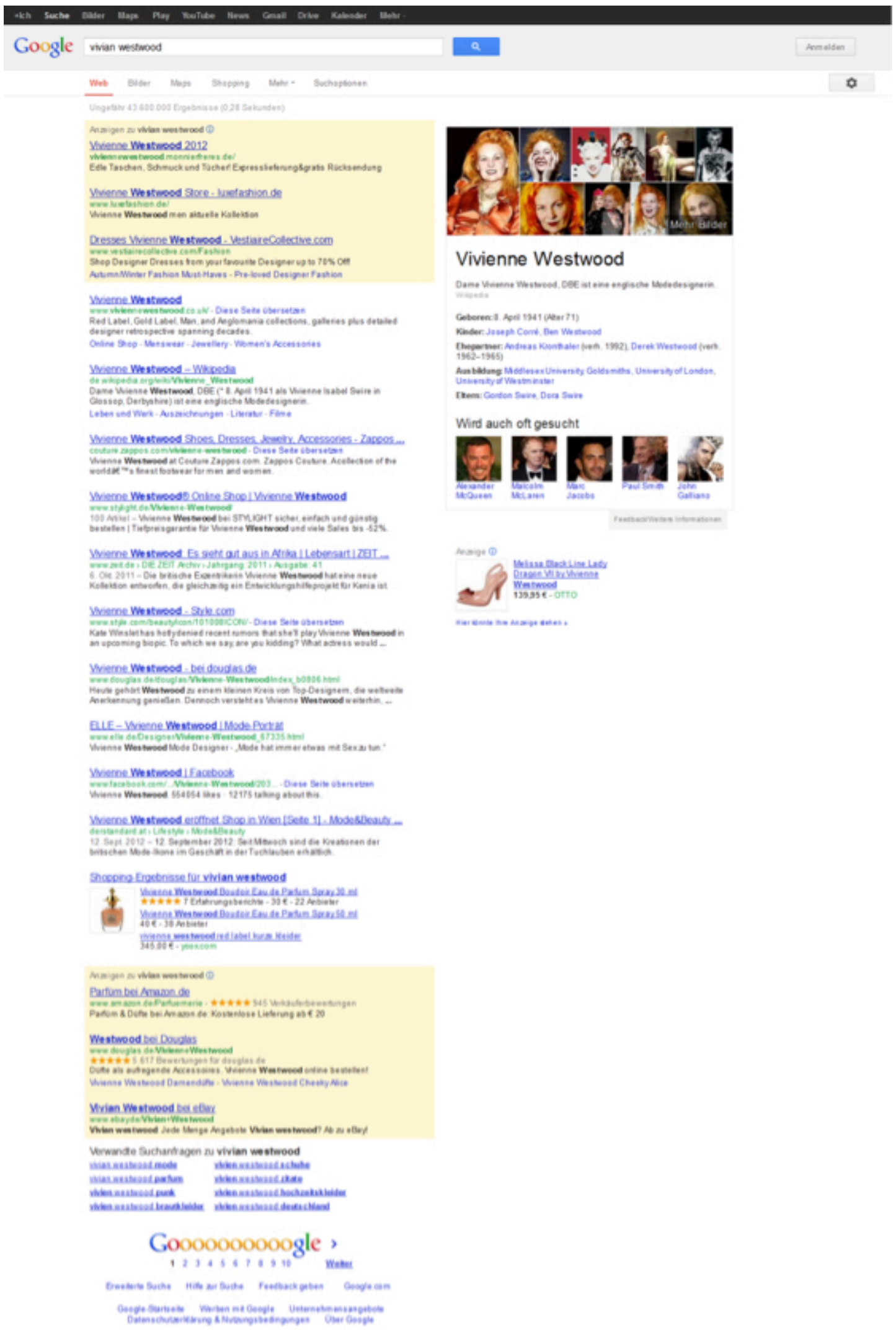

*Figure 7*. **Screenshot provided for the "Vivienne Westwood" task (Q10)**

**Table 7.** ***Results for Vivienne Westwood task (question 10) by self-assessments of the participants regarding search skills***

| | All users | German grade 1 (n = 454) % | German grade 2 (n = 454) % | German grade 3 (n = 81) % | German grade 4 (n = 9) % | German grade 5 (n = 1) % | German grade 6 (n = 1) % |
|---|---|---|---|---|---|---|---|
| ALL advertisements marked correctly | 15.8 | 15.4 | 16.5 | 14.8 | 0.0 | 0.0 | 100.0 |
| Advertisements only (but did not identify all advertisements) | 44.9 | 47.1 | 45.2 | 33.3 | 22.2 | 0.0 | 100.0 |
| Organic results marked as advertisements | 31.9 | 28.0 * | 33.3 * | 44.4 * | 55.6 * | 0.0 | 0.0 |
| Universal Search results marked | 14.6 | 15.4 | 13.2 | 16.0 | 22.2 | 100.0 | 0.0 |
| All search results marked as ads | 0.8 | 0.9 * | 0.7 * | 0.0 * | 11.1 * | 0.0 * | 0.0 * |
| Marked only ads on the right-hand side | 14.3 | 15.2 | 14.5 | 9.9 | 0.0 | 0.0 | 0.0 |
| Marked only ads above the organic results | 12.9 | 11.2 | 13.7 | 18.5 | 11.1 | 0.0 | 0.0 |

*Note.* *p < 0.05. For German grades, 1 is the best possible grade and 6 the worst.

**Table 8.** ***Results for Vivienne Westwood task (question 10) by self-reported knowledge about Google's business model; knowledge of placing ads on the SERPs; and knowledge of distinction between advertisements and organic results***

| | **Knowledge of Google's business model** | | | | **Knowledge of paying Google to place ads on the SERPs** | | | **Knowledge of distinction between advertisements and organic results**** | | |
|---|---|---|---|---|---|---|---|---|---|---|
| | correct (n = 606) | incorrect (n = 93) | don't know (n = 95) | incorrect statement (n = 206) | yes (n = 733) | no (n = 64) | don't know (n = 203) | yes (n = 425) | no (n = 195) | don't know (n = 113) |
| | % | % | % | % | % | % | % | % | % | % |
| ALL advertisements marked correctly | 18 * | 5.4 * | 4.3 * | 19.4 * | 16.9 | 15.6 | 11.8 | 21.4 * | 7.7 * | 15.9 * |
| Advertisements only (but did not identify all advertisements) | 49.0 * | 31.2 * | 31.6 * | 45.1 * | 46.2 * | 53.1 * | 37.4 * | 53.2 * | 32.3 * | 44.2 * |
| Organic results marked as advertisements | 30.4 | 41.9 | 26.3 | 34.5 | 32.9 | 20.3 | 32.0 | 28.9 * | 44.6 * | 27.4 * |
| Universal Search results marked | 20.4 | 12.9 | 15.8 | 16.5 | 15.4 | 6.3 | 14.3 | 12.7 * | 21.5 * | 15.0 * |
| All search results marked as ads | 1.0 | 2.2 | 0.0 | 0.0 | 1.0 | 0.0 | 0.5 | 0.2 * | 3.1 * | 0.0 * |
| Marked only ads on the right-hand side | 14.2 | 10.8 | 15.8 | 15.5 | 11.6 * | 20.3 * | 22.2 * | 8.0 * | 19.0 * | 12.4 * |
| Marked only ads above the organic results | 14.2 | 15.1 | 4.2 | 12.1 | 13.6 | 12.5 | 10.3 | 13.6 | 14.4 | 12.4 |

*Note.* *p < 0.05, ** *FILTER: only participants answered the former question with "yes". (n=733)*

In question 11, users were shown a SERP for the query "buy laptop" and were asked to mark all results that Google was not paid for. In this case, advertisements were shown on top of the results list and also on the right-hand side. Here, there were two different types of advertising: text-based ads (as in the other tasks) as well as ads with product images. Furthermore, within the list of organic results, there was a box showing results from Google Shopping, which at that time was showing organic results from a product database (see Figure 8).

We can see that users clearly were confused by the search results presentation: Only 7.6% were able to mark all organic results correctly (including the Universal Search box showing the shopping results). 40.2% percent marked only organic results, but not the complete set. 30.7% (correctly) marked the Shopping Search box as organic. However, 29.4% marked at least one advertisement as an organic result.

Significant differences were found for the Google Shopping box; for completing the entire task correctly; and for marking only organic results (see Table 10).

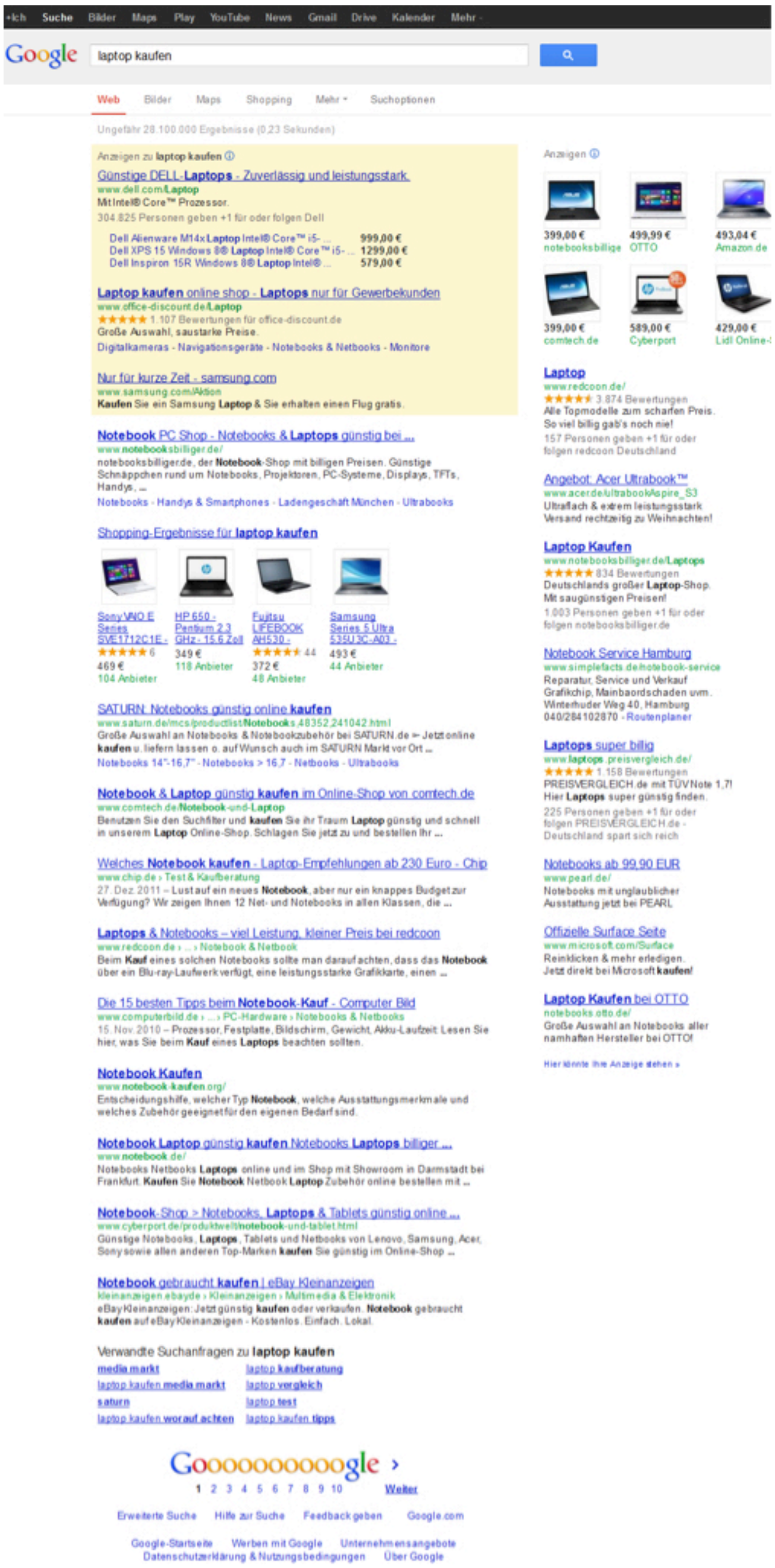

***Figure 8.*** **Screenshot displayed for the "buy laptop" task (Q11)**

**Table 9.** ***Results for the "buy laptop" task (question 11) by self-assessment of the participants regarding search skills***

| | All users | German grade 1 (n = 454) | German grade 2 (n = 454) | German grade 3 (n = 81) | German grade 4 (n = 9) | German grade 5 (n = 1) | German grade 6 (n = 1) |
|---|---|---|---|---|---|---|---|
| | | % | % | % | % | % | % |
| ALL organic results and Universal Search results marked correctly | 7.6 | 7.3 | 7.5 | 9.9 | 11.1 | 0.0 | 0.0 |
| Organic results and Universal Search results only (but did not mark all results correctly) | 40.2 | 38.1 | 40.5 | 48.1 | 55.6 | 100.0 | 0.0 |
| Universal Search results marked | 30.7 | 30.6 | 30.8 | 29.6 | 44.4 | 0.0 | 0.0 |
| Advertisements marked as organic results | 29.4 | 32.8 | 27.8 | 19.8 | 22.2 | 0.0 | 100.0 |
| All search results marked as organic results | 1.0 | 1.1 | 1.1 | 0.0 | 0.0 | 0.0 | 0.0 |

*Note.* * $p < 0.05$. For German grades, 1 is the best possible grade and 6 the worst.

**Table 10.** ***Results for the "buy laptop" task (question 11) by self-reported knowledge about Google's business model; by knowledge of placing ads on the SERPs; and by knowledge of distinction between advertisements and organic results***

| | **Knowledge of Google's business model** | | | | **Knowledge of paying Google to place ads on the SERPs** | | | **Knowledge of distinction between advertisements and organic results**** | | |
|---|---|---|---|---|---|---|---|---|---|---|
| | correct (n = 606) | incorrect (n = 93) | don't know (n = 95) | incorrect statement (n = 206) | yes (n = 733) | no (n = 64) | don't know (n = 203) | yes (n = 425) | no (n = 195) | don't know (n = 113) |
| | % | % | % | % | % | % | % | % | % | % |
| ALL organic results and Universal Search results marked correctly | 9.2 | 3.2 | 4.2 | 6.3 | 8.6 | 4.7 | 4.9 | 12.0 * | 3.6 * | 4.4 * |
| Organic results and Universal Search results only (but did not identify all results ) | 39.1 | 39.8 | 38.9 | 44.2 | 42.7 * | 31.3 * | 34 * | 45.6 | 39.0 | 38.1 |
| Universal Search results marked | 33.5 * | 26.9 * | 17.9 * | 30.1 * | 32.5 * | 35.9 * | 22.7 * | 34.6 | 27.2 | 33.6 |
| Advertisements marked as organic results | 30.9 | 25.8 | 26.3 | 28.2 | 28.9 | 29.7 | 31.0 | 27.3 | 29.7 | 33.6 |
| All search results marked as organic results | 1.0 | 1.1 | 2.1 | 0.5 | 0.5 * | 3.1 * | 2 * | 0.2 | 1.0 | 0.9 |

*Note.* *$p < 0.05$, ** *FILTER: only participants answered the former question with “yes”. (n=733)*

## DISCUSSION AND CONCLUSION

If we consider the results from tasks 7 to 11 together (table 11), we can see that only 1.3 percent of participants were able to identify all results correctly. Even if we only consider those users whose identifications were all correct (but who did not mark *all* results that should have been marked; 9.6%), we find that only 10.9% of users made no incorrect identifications. This leads us to the conclusion that even when users are aware of the distinction between ads and organic results, they still have difficulties in many instances.

3.9% of participants marked at least one Universal Search result as an ad, which may be due to some confusion regarding the graphical display of search results.

It is surprising that only a small proportion of users assessed whether a result should be considered an ad or an organic result solely on the basis of its position. Just 1.5% marked only results on the right-hand side of the SERP as ads, and just 0.7% marked only ads at the top the SERP as ads (but not on the right-hand side).

Taking these results together, we can see that a large percentage of users is not able to reliably distinguish between organic results and ads on Google’s SERPs (RQ1). Regarding differences between user groups (RQ1a-RQ1c), we did not find significant differences concerning self-assessment, knowledge about Google’s methods of generating revenue, nor users’ assessment of whether it is possible to distinguish between organic results and paid advertisements. Furthermore, many users are not aware of how search engine companies make money (RQ2).

**Table 11.** ***Summary of results across all tasks (Q7-Q11) for all participants***

| | Total (n = 1,000) % |
|---|---|
| ALL areas marked correctly | 1.3 |
| All identifications correct (but not all areas identified) | 9.6 |
| Incorrect assignments (at least one organic result labelled as advertisements or vice versa) | 0.6 |
| All search results marked as advertisements or as organic results | 0.1 |
| Universal Search results marked | 3.9 |
| Marked only ads on the right-hand side | 1.5 |
| Marked only ads above the organic results | 0.7 |

Our results are in line with those from previous, non-representative industry studies (Bundesverband Digitale Wirtschaft, 2009; Charlton, 2013; Wall, 2012). Our study demonstrates that German Google users experience considerable difficulties in distinguishing between paid advertisements and organic results on the results pages. As search engines are a major tool not only for finding information on the Web, but also more generally for acquiring knowledge to base decisions upon, our results are also significant with respect to the broader question of how users are influenced by interest-driven communications when they are searching for information. When users trust a search engine, but fail to recognize advertisements, the trust they place in the search engine may carry over to the results generated through advertising. This leads to the larger question of how we should regard context-based advertising in search engines — are they merely ads comparable to other forms of advertisements, or should we regard them as a results type, or even as results that should be treated the same way we treat organic results?

While we did not investigate the effect of alternative labelling on users' ability to distinguish between the two results types, we can at least assume that a clearer distinction on the SERPs would lead to users performing better when selecting either organic results or ads. In this regard, regulating bodies should consider asking for a clearer distinction on the SERPs. Past attempts (e.g., from the Federal Trade Commission in the U.S.) may not have had the desired effect.

As search engines in general, and Google in particular, are used for a wide variety of information needs, it is important that users are aware of the results type they select. However, it seems that search engines like Google at best tolerate the fact that users are being misguided into clicking on advertisements when they think they are clicking on an organic result.

To our knowledge, ours is the first study using a representative sample to find out how well users are able to distinguish between ads and organic results on the SERPs. While we surely only took a sample of the German online population, and therefore can only make statements about that group, our study could be seen as a example of using representative samples in information behaviour research. Furthermore, our study is not merely based on *asking* users how to distinguish between the two results types, but actually letting them label the results they think belong to a certain type.

Results presentation in search engines constantly changes. For instance, since our data was collected, Google changed the ads labels (from shading and labelling to a new label; see Marvin, 2016), and dropped the right-hand side ads entirely.

Future research is needed not only on the effect of different types of results labelling on user selection behaviour, but also on the influence of the actual relevance of the ads. One could ask whether users who are able to distinguish between organic results and ads are more successful when it comes to search tasks for which the presentation of results is influenced by advertisers.

## ACKNOWLEDGEMENTS

No external funding was provided to conduct this study.